 \documentclass[final,5p,times,twocolumn,sort&compress]{elsarticle}

\usepackage{amssymb}
\usepackage[utf8]{inputenc}
\usepackage[]{amsmath}
\usepackage[]{graphicx}
\journal{Physics Letters B}

\begin{document}

\begin{frontmatter}



\title{Is the unusual near-threshold potential behavior in elastic scattering of weakly-bound nuclei a precision error? }

\author[label1,label2]{Rodrigo Navarro P\'erez}
\ead{rnavarroperez@sdsu.edu}
%\email[]{rnavarroperez@sdsu.edu}
\author[label2]{Jin Lei}
\ead{jinl@ohio.edu}
%\email[]{jinl@ohio.edu}
%\thanks{}
\address[label1]{Department of Physics. San Diego State University. 5500 Campanile Drive, San Diego, California 02182-1233, USA}
\address[label2]{Institute of Nuclear and Particle Physics, and Department of Physics and Astronomy, Ohio University, Athens, Ohio 45701, USA}


%\author{}
%
%\address{}

\begin{abstract}
We present the first example of a rigorous uncertainty quantification on elastic 
Nucleus-Nucleus scattering at energies near the Coulomb barrier. Experimental 
data has been analyzed using an energy-dependent effective optical model potential 
with physical constraints imposed. We confirm the compatibility of these 
uncertainties with the well known Coulomb threshold anomaly, explained in 
terms of a dispersive relation, and contrast our results with previous analyses 
that suggest otherwise.
\end{abstract}

\begin{keyword}
nuclear reaction \sep dispersive relation \sep optical potential \sep uncertainty quantification \sep Coulomb threshold anomaly \sep statistical analysis
%%% keywords here, in the form: keyword \sep keyword
%
%%% PACS codes here, in the form: \PACS code \sep code
%
%%% MSC codes here, in the form: \MSC code \sep code
%%% or \MSC[2008] code \sep code (2000 is the default)
%
\end{keyword}

\end{frontmatter}


\section{Introduction}
The reactions of weakly-bound stable and unstable nuclei have been extensively investigated for several decades. Special attention has been devoted to reactions at low energies close to the Coulomb barrier, where these reactions are primarily dominated by fusion, and direct reactions in which non-elastic scattering occurs. 
The energy dependence of the effective nuclear potential between projectile and target around Coulomb barrier is an important information, which has been widely studied theoretically and experimentally. 

From a theoretical point of view, the nuclear part of the optical model potential (OMP), can be written as $U(E)=V(E)+iW(E)$. Causality considerations and microscopic theory relate the energy dependence of the real and imaginary part of the OMP through a dispersive relation that results in $V(E)=V_0(E)+\Delta V(E)$. The term $V_0(E)$ exhibits a slow and smooth dependence on the energy, while the term $\Delta V(E)$  depends upon $W(E')$ at all energies $E'$,
\begin{equation}
\Delta V(E) = \frac{\mathcal{P}}{\pi} \int^{\infty}_0 \frac{W(E')}{E'-E}dE',
\label{eq:dispersion}
\end{equation}
where $\mathcal{P}$ is Cauchy principal value. This relation was first introduced by Feshbach~\cite{Fes58,Fes62} from a microscopic point of view and later investigated by Cornwall and Ruderman~\cite{Cornwall1962} through causality relations.

%It can be seen that any localized, rapid variation of the imaginary potential $W$ must be accompanied by a similarly localized variation in the real term $V$, this is referred as the ``anomaly'' mentioned above. 

On the other hand, the energy dependence of the optical potential was first experimentally found by Lipperheide and Schmidt~\cite{LIPPERHEIDE196865}. Later in  mid-1980s~\cite{Baeza84,Lilley85,Fulton85} a strong energy dependence of the potential at energies close the Coulomb barrier was discovered. The imaginary potential decreases rapidly with the effective closure of the nonelastic channels
when the energy is reduced to below the barrier and the real potential shows an ``anomalous'' variation around the barrier~\cite{Nagarajan85,Satchler91,Mahaux86}. 
Therefore, the barrier acts as a natural ``threshold'' for processes involving the action of the nuclear force. This behavior is commonly referred to as the Threshold Anomaly (TA).

However, it was later suggested that the near-threshold potential behavior for weakly-bound nuclei, such as $^6$Li, may be different compared to tightly-bound nuclei. The imaginary part of the effective potential increases or remains constant as the bombarding energy decreases towards the Coulomb barrier (see Ref.~\cite{Hussein06,li6al27,li6si28,li6ni58,li6co59,li6zn64,li6se80,li6zr90,li6sn112,li6ba138,li6sm144,li6pb208,li6bi209,li6th232,li7al27,li7Zn64,li7pb208,be9al27,be9y89,be9sm144,be9pb208}), an effect which has been termed as Breakup Threshold Anomaly (BTA) . Recently, an abnormal threshold anomaly was reported to be found for the exotic nuclear system $^6$He + $^{209}$Bi~\cite{Yang17}, in which the authors concluded that the dispersion relation of Eq.(\ref{eq:dispersion}) is not applicable for this system.

For a heavy-ion nuclear reaction, the OMP parameters are usually extracted by fitting  experimental elastic scattering data. At energies close to and below the Coulomb barrier, the elastic scattering cross section is close to the Rutherford cross section, therefore the nuclear part of the OMP is mainly hidden. For reactions induced by unstable nuclei, the situation becomes even worse due to technical limitations in the intensity and/or the phase-space qualities of radioactive ion beams. In view of these facts, large uncertainties are introduced when extracting the parameters of OMP by directly fitting the data. Furthermore, the analyses presented in Refs.~\cite{Yang17,yang17prc,li7Zn64,li6ni58,li6zr90} may be over-fitted to the data as suggested by their $\chi^2/{\rm d.o.f.} \ll 1$. In order to reduce the effect of these large uncertainties on the determination of the OMP parameters, we propose that physical constraints be imposed on the model used to analyze the data. 

In addition, it should be noted that most previous a\-na\-ly\-ses may be underestimating the error bar in the OMP parameters in two ways. First, the OMP parameters at one energy are determined independently from all other energies, that is a fit is made for each energy at which experimental data is available keeping any common parameters fixed. Second, on each fit a linear correlation between the fitting parameters is assumed. In fact a large number of optical potentials are adjusted using codes relying on the MINUIT package of function minimization and error analysis. As stated on the corresponding manual ``errors based on the Minuit error matrix take account of all the parameter correlations, but not the non-linearities''~\cite{james1972}. However, due to the dispersive relation of Eq.(\ref{eq:dispersion}), we expect the correlations between the coefficients of the real and imaginary parts to have significant non-linearities specially between different energies. To avoid underestimating the uncertainties we perform a fit of all energy dependent and independent parameters simultaneously to data at all available energies and do not assume linear correlations by employing a Monte-Carlo technique as described below.

In this Letter we present a proper extraction of the OMP parameters for weakly-bound nuclei induced reaction. As a result the OMP energy dependence is found to exhibit the usual near-threshold potential behavior (i.e. TA) and the model obtained describes the experimental with $\chi^2/{\rm d.o.f.} \simeq 1$, indicating agreement between theory and experiment within experimental uncertainties.

\section{Data analysis}
For the purposes of this work we analyze elastic scattering data from two weakly-bound systems, $^6$Li + $^{209}$Bi~\cite{li6bi209} and $^6$He + $^{208}$Pb~\cite{SanchezBenitez:2005jj,Sanchez-Benitez:2008krm,Marquinez-Duran:2012ilt,Kakuee:2005kq}. In both cases we use an OMP to describe the nuclear part of the Nucleus-Nucleus interaction. The real and imaginary parts of the OMP consist of a Woods-Saxon potential with energy independent range and diffuseness parameters along with energy dependent strength coefficients, i.e.\footnote{For simplicity, we assume potentials are local and  only consider central parts. }
\begin{eqnarray}
	V(r,E) &=& -\frac{V_E}{1+e^{(r-R_V)/a_V}}, \\
	W(r,E) &=& -\frac{W_E}{1+e^{(r-R_W)/a_W}}.
\end{eqnarray}
%where $R_i = r_i(A_T^{1/3} + A_P^{1/3})$ and $A_T$ and $A_P$ correspond to the number of nucleons in the projectile and target respectively.
The scattering amplitude was calculated by solving the Schr\"odinger equation using the modified Numerov method outlined in Ref.~\cite{thompson2009nuclear}. The results were verified for accuracy by comparing with calculations made with the computer code FRESCO~\cite{fresco}.

In a first step the parameters in the OMP were adjusted using the usual least squares procedure. In particular, the $\chi^2$ function was minimized using the Levenberg-Marquardt method outlined in~\cite{press1992numerical}. An advantage of this method is that a first order approximation to the Hessian matrix is obtained which allows estimating the parameter uncertainty through the co-variance matrix, assuming a multivariate normal distribution. In order to ensure a sound quantification of uncertainties we applied the self-consistent $3\sigma$ criterion to the experimental data in order to identify and exclude statistically inconsistent data points. This method has been successfully used in the context of analyzing $pp$, $np$ and $\pi\pi$ scattering data~\cite{Perez:2013jpa,Perez:2015pea}. 

\begin{figure*}
\centering
\includegraphics[width=\linewidth]{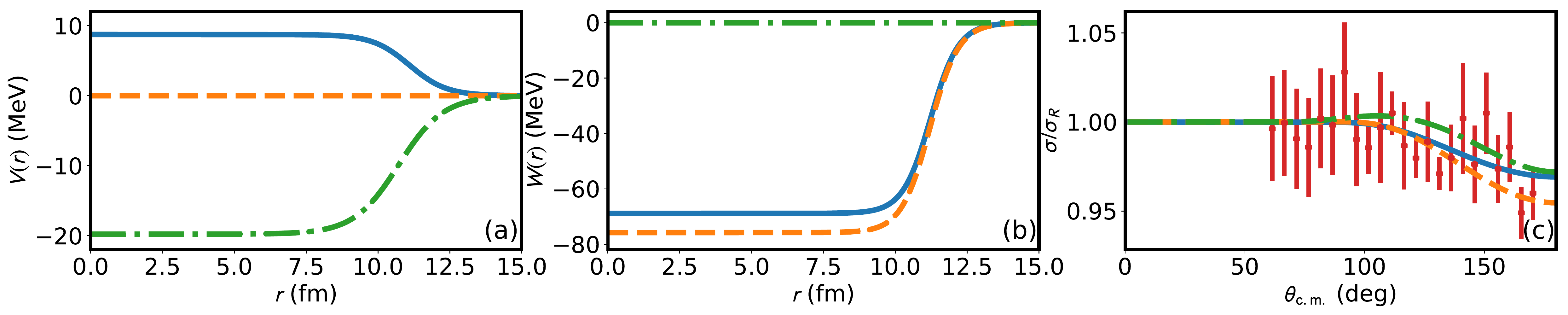}%
\caption{\label{fig:low_energy_scattering}(Color on-line) Three completely different OMPs describing the $^6$Li + $^{209}$Bi elastic scattering at $E_{\rm LAB}=24$ MeV. The left and middle panels show the real and imaginary part of the OMP, respectively. The coulomb contribution is not included in the figures. The right panel shows the corresponding elastic scattering (matched by color and line type) compared to experimental data~\cite{li6bi209} (red error bars).}
\end{figure*}

After the initial $\chi^2$ minimization the obtained co-variance matrix indicates that, as expected, statistical uncertainties in the OMP parameters are rather large in particular at low energies. These large uncertainties are reflected in the calculated scattering amplitude being rather insensitive to the OMP parameters. A clear example of this insensitivity can be seen in Fig.~\ref{fig:low_energy_scattering}. Three substantially different OMPs for the $^6$Li + $^{209}$Bi reaction at $E_{\rm LAB} = 24$ MeV result in very similar cross sections, all of them compatible with experimental data. Notice that the OMP represented by blue solid line even has a repulsive real part, instead of the expected attractive interaction. To deal with these large uncertainties we impose additional constraints to the OMP parameters based on physical principles. In a second least squares analysis of the data we simply impose that the OMP must be an attractive interaction (i.e. that $V_E$ and $W_E$ are positive numbers). For a third least squares analysis we impose the additional constraint of the strength coefficients at energies below the Coulomb barrier being small compared to the strength at higher energies. The reasoning of this being that, like in the case of tightly-bound nuclei, as the energy goes below the Coulomb barrier there is no need for the imaginary part of the OMP to absorb the inelastic channels that are already closed. To impose small strength coefficients at low energy we modify the usual $\chi^2$ by adding terms that penalize large values of the corresponding parameters. These penalty terms can be expressed as
\begin{equation}
\tilde{\chi}^2 = \chi^2 + \frac{(p-p_0)^2}{(\Delta p)^2},
\label{eq:priors}
\end{equation}
where $p$ is one of the OMP parameters, $p_0$ is some small central value and $\Delta p$ is the expected variation range for $p$. The effect of adding this penalty function is that in order to minimize the modified $\tilde{\chi}^2$, the parameter $p$ needs to be close to the small central value $p_0$. While imposing such penalty function might seem completely arbitrary, this type of constraint is the frequentist equivalent of using a Gaussian prior in Bayesian statistics. The use of priors is a common and justified practice in Bayesian parameter estimation~\cite{Furnstahl:2014xsa}. 

To properly quantify the statistical uncertainty that propagates from the experimental data to the OMP parameters through the $\chi^2$ minimization we make use of simple a Monte-Carlo technique. For every experimental data point $O_i$ with error bar $\sigma_i$ we make the substitution $O_i \rightarrow O_i + \mathcal{N}(0,\sigma_i)$, where $\mathcal{N}(0,\sigma_i)$ are random variates with a normal distribution, and readjust the OMP parameters by minimizing the resulting $\chi^2$ function. We repeat this process until we have a sufficiently large number of samples for the OMP parameters\footnote{In practice we obtain 1000 samples, although a smaller number might already give a robust estimation of the parameters distribution}. This method relies on the assumption that the residuals in the $\chi^2$ function follow the standard normal distribution; this assumption has been stringently verified for every least squares analysis on this work using the hypothesis testing method described in~\cite{Perez:2014kpa}. As mentioned earlier, we opt for this method of uncertainty quantification over the usual co-variance matrix since the latter assumes linear correlations between the OMP parameters but from Eq.(\ref{eq:dispersion}) we in fact expect non linear correlations which could result in underestimated error bars. Furthermore, the co-variance matrix as a representation of the uncertainties and correlations between the model parameters assumes small errors from the start, which Fig.~\ref{fig:low_energy_scattering} already shows not to be the case. The Monte-Carlo method automatically extracts a sample of the parameters distribution without assuming any kind of theoretical distribution for them. The same method, sometimes called parametric bootstrap~\cite{Pastore:2018xuu}, has been used successfully to estimate uncertainties in \emph{ab initio} calculations of the binding energy of light nuclei~\cite{Navarro:2014laa, Perez:2014jsa, Perez:2015bqa} and in the analysis of Compton scattering data~\cite{Pasquini:2017ehj}.

\section{Results and discussion}
\begin{figure*}
\centering
\includegraphics[width=\linewidth]{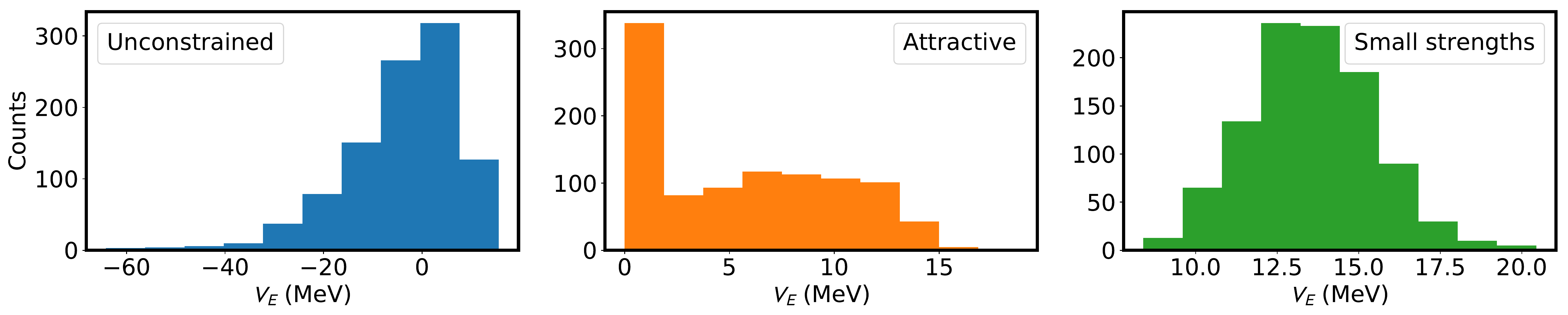}%
\caption{\label{fig:histograms}(Color on-line) Histograms showing the distribution of the strength parameter $V_E$ at $E=26$~MeV resulting from the Monte-Carlo sampling analyzing $^6$Li + $^{206}$Bi elastic scattering data. Three different analysis are presented, a completely unconstrained fit (left panel), a fit imposing only attractive optical potentials (middle panel), and a fit constraining the strength coefficients $W_E$ at energies below the Coulomb barrier to be small compared to the strength at higher energies (right panel).}
\end{figure*}

\begin{figure*}
\centering
\includegraphics[width=0.5\linewidth]{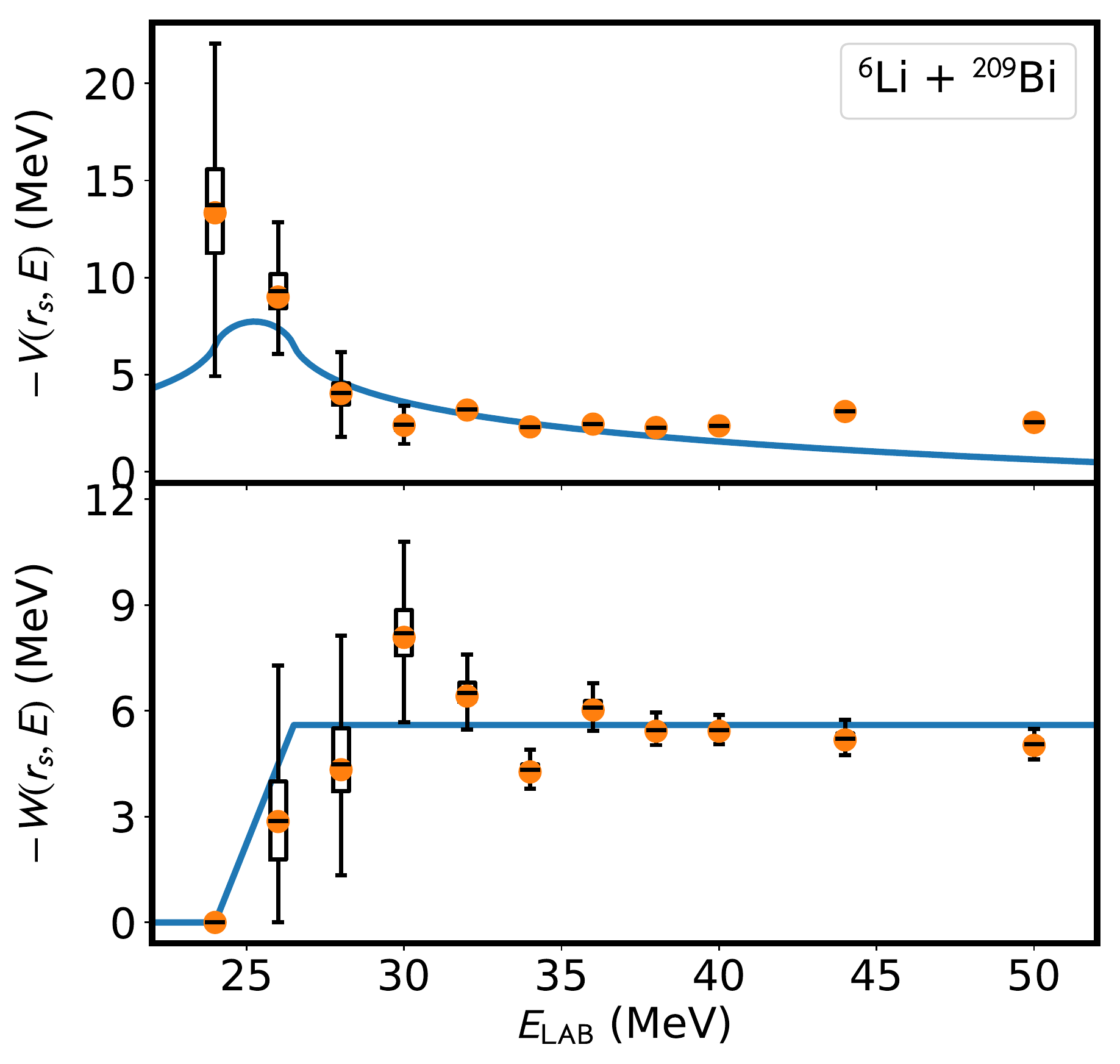}%
\includegraphics[width=0.5\linewidth]{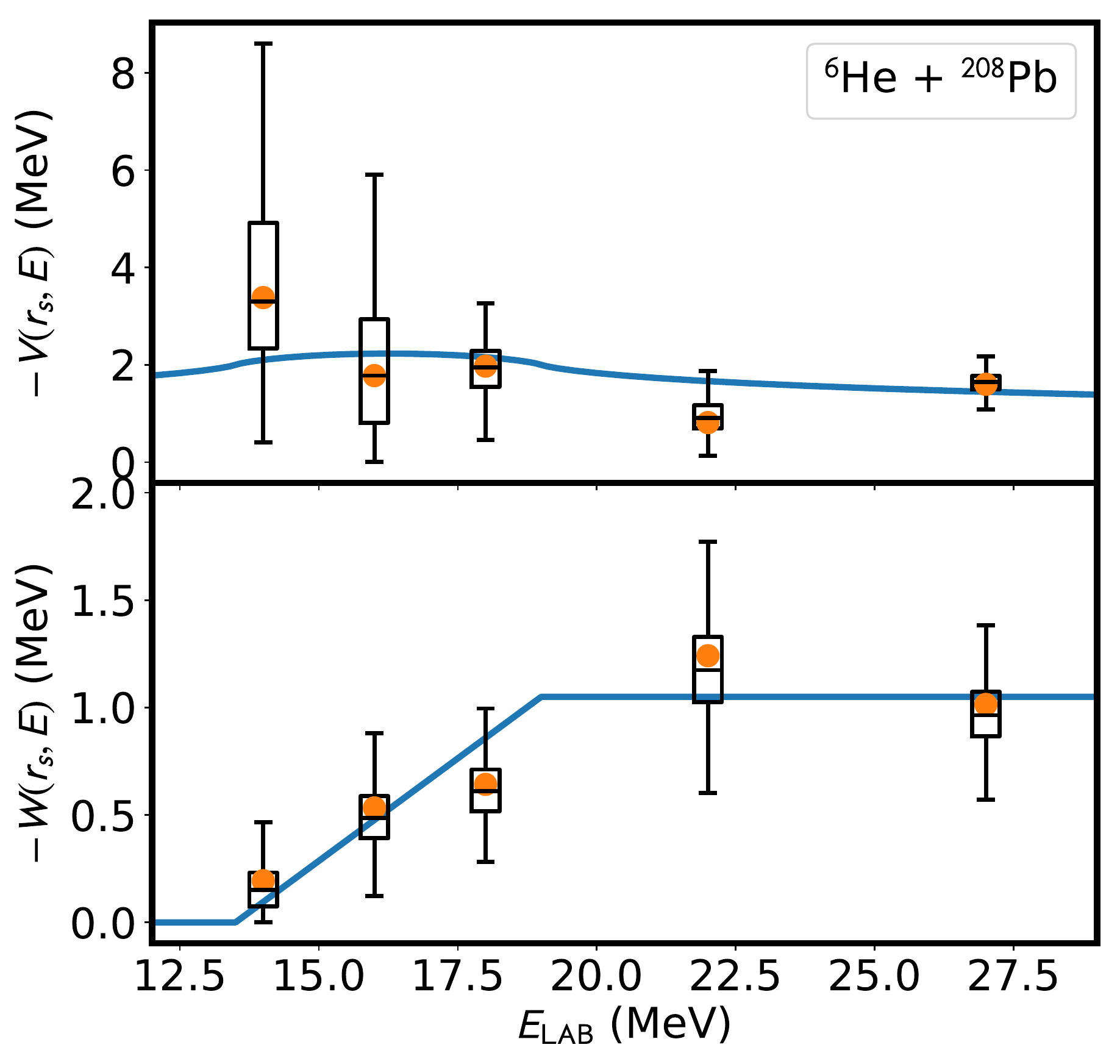}%
\caption{\label{fig:dispersion_Li6_Bi209} (Color on-line) Energy dependence of the real (top panels) and imaginary part (bottom panels) of the OMPs at the sensitivity radius $r_s = 10.15$~fm for the $^6$Li + $^{209}$Bi reaction (left panels) and $r_s = 10.6$~fm for the $^6$He + $^{208}$Pb reaction (right panels). The box and whiskers plots represent the samples obtained from the Monte-Carlo technique (as described in the text). The orange points are determined by the central values of the parameters. The blue solid line for the imaginary part is a linear approximation to the energy dependence while the blue solid line for the real part is obtained from the dispersive relation in Eq(\ref{eq:dispersion}).}
\end{figure*}

Our results are summarized in tables~\ref{tab:chi_square} and~\ref{tab:dispersion parameters}, along with figures~\ref{fig:histograms} and~\ref{fig:dispersion_Li6_Bi209}. Table~\ref{tab:chi_square} shows the $\chi^2$ per number of degrees of freedom for each of the three OMP fits to the $^6$Li+$^{209}$Bi and $^6$He+$^{208}$Pb experimental scattering data. For the interested reader, an additional table in the supplementary material gives the resulting parameters of the potential constrained to have small strengths coefficients in the imaginary part at energies below the Coulomb barrier. Given the number of data and parameters in each reaction, the total $\chi^2/{\rm d.o.f.}$ obtained agrees with the expected value of $1$ at a level between two and three sigmas. Furthermore, the values in the table show that imposing the physical restrictions of attractive OMP first, and small strengths at energies below the Coulomb barrier later, has little to no effect on the average discrepancies between theoretical model and experimental data. That is, the three types of analyses reproduce the experimental data equally well, as expected from the three vastly different potentials in figure~\ref{fig:low_energy_scattering}. However, the imposition of constraints based on physical expectations of the OMP does have an effect on the uncertainty of the potential parameters by reducing the region in hyper-parameter space that can be explored when the Monte-Carlo analysis is performed. This effect is clearly reflected in Fig.~\ref{fig:histograms} which shows a histogram for the strength parameter $V_E$ at $E=26.0$~MeV, resulting from the Monte-Carlo sampling analyzing the $^6$Li + $^{209}$Bi data. The resulting parameters in the unconstrained analysis cover a large range of values, most of them corresponding to a repulsive potential. As physical constraints are imposed on the analysis the range covered by the resulting parameters gets reduced. It is noteworthy that the constraint of small strength coefficients is imposed on the \emph{imaginary} part of the potential only, yet this has a clear effect of constraining the \emph{real} part as well.

\begin{table} % add [H] placement to break table across pages
\centering
 \caption{\label{tab:chi_square} $\chi^2$ per number of degrees of freedom after adjusting the OMP parameters to two different Nucleus-Nucleus elastic scattering reactions. For each reaction three different types of analyses are made. A completely unconstrained fit, a fit where the OMP is enforced to be attractive and a fit where the strength coefficients at energies below the Coulomb barrier are made to be small via penalty functions like the one shown in Eq.(\ref{eq:priors})}
% \begin{ruledtabular}
  \begin{tabular}{l c c c}
  \hline\hline
   & \multicolumn{2}{c}{$\chi^2$/{\rm d.o.f.}} \\
   \hline
   Reaction      & $^6$Li + $^{209}$Bi & $^6$He + $^{208}$Pb \\
   Number of data    & 227 &  246 \\
   \hline
   Unconstrained   & 1.32 & 0.79 \\
   Attractive OMP  & 1.35 & 0.79 \\
   Small strengths & 1.38 & 0.80 \\
   \hline\hline
  \end{tabular}
% \end{ruledtabular}
\end{table}

For each reaction, the result of the Monte-Carlo a\-na\-ly\-sis described earlier is a collection of 1000 OMPs with the physical constraint of small strengths at energies below the Coulomb barrier imposed. While the parameters obtained differ from one fit to another, the description of the experimental data by every OMP is statistically equivalent. This collection of potentials allows identifying the radius at which the potential changes the least, also known as the sensitivity radius $r_s$. As it is customary, we present the energy dependence of the OMP at the sensitivity radius, selecting $r_s=10.15$~fm for the $^6$Li + $^{209}$Bi reaction and $r_s = 10.6$~fm for the $^6$He + $^{208}$Pb reaction\footnote{Of course the position of $r_s$ can be different for the real and imaginary part of the potential and for each energy. In practice we take the average of all the sensitivity radii, which are mostly spread between 8 and 12 fm. For reassurance we checked that the same type of energy dependence can be observed at different values of $r_s$}. Fig.~\ref{fig:dispersion_Li6_Bi209} shows the results of the Monte-Carlo analysis in the form of box and whisker plots. The size of the box represents the inter-quartile range ($IQR$) of the 1000 OMPs, the horizontal line indicates the position of the median and the end points of the whiskers give the position of the first element in the sample that has a distance to the end of the box of least $1.5 IQR$. Elements farther away from the endpoints are called flyers and are not shown in these figures. We opt for this graphical representation of the data, in lieu of the usual error bar, to emphasize the sometimes asymmetric nature of the distribution of the resulting sample. For reference we also show, as an orange dot, the potential obtained from the $\chi^2$ minimization.

The energy dependence of the imaginary part of the OMP is parametrized with a linear approximation at low energies. Following the prescription outlined in section 3.4 of Ref.~\cite{Mahaux86} we assume that $|W(E)|$ rises from zero, at an energy $E_1$, to a certain value $W_0$, at another energy $E_2$, to then remain constant. This linear dependence is shown in Fig.~\ref{fig:dispersion_Li6_Bi209} as a blue solid line in the bottom panels. Using Eq. (\ref{eq:dispersion}) one can calculate the energy dependence of the real part which can then be added to a constant value $V_0$. The actual calculation was done using the analytic expression
\begin{equation}
\Delta V(E) = \frac{W_0}{\pi} \left(\epsilon_1  \ln |\epsilon_1| - \epsilon_2  \ln |\epsilon_2|\right),
\end{equation}
where $\epsilon_i = (E-E_i)/(E_2-E_1)$. For a derivation of this expression see section 3.4 of Ref.~\cite{Mahaux86}. The result from the dispersion relation is shown as a blue solid line in the top panels. The values for $W_0$, $V_0$, $E_1$, and $E_2$, which were chosen to show agreement between the dispersion relation and the Monte-Carlo analysis presented as the box and whiskers plots, are presented in table~\ref{tab:dispersion parameters}.

% \begin{figure}
% \includegraphics[width=\linewidth]{fig_dispersion_6He208Pb_prior.pdf}%
% \caption{\label{fig:dispersion_He6_Pb208} Same as figure~\ref{fig:dispersion_Li6_Bi209} at the sensitivity radius $r_s = 10.6$~fm and the $^6$He + $^{208}$Pb reaction}
% \end{figure}

\begin{table} % add [H] placement to break table across pages
\centering
 \caption{\label{tab:dispersion parameters} Parameters necessary for the calculation of the energy dependence $\Delta V(E)$ in the real part of the optical potential as outlined in section 3.4 of Ref.~\cite{Mahaux86}. All values are in units of MeV}
% \begin{ruledtabular}
  \begin{tabular}{l c c c}
  \hline\hline
   Parameter      & $^6$Li + $^{209}$Bi & $^6$He + $^{208}$Pb \\
   \hline
   $V_0$ & 6.5  & 2.00 \\
   $W_0$ & 5.6  & 1.05 \\
   $E_1$ & 24.0 & 13.5 \\
   $E_2$ & 26.5 & 19.0 \\
   \hline\hline
  \end{tabular}
% \end{ruledtabular}
\end{table}

It should be mentioned that the main reason for the agreement with the dispersive relation is the imposition physical constraints, in this case small strength coefficients at energies below the Coulomb barrier as shown in Eq.~(\ref{eq:priors}). While the use of Monte-Carlo techniques allow for a rigorous and sound propagation of the experimental uncertainty into the OMP parameters, additional physical information is required for the imaginary part of the potential to decrease with decreasing energy. Previous analyses of elastic Nucleus Nucleus scattering with weakly-bound nucleus have used Monte-Carlo methods to calculate the uncertainty in OMP parameters but found similar results to analyses with the usual co-variance matrix~\cite{Abriola:2017, Abriola:2015ufa}. Such analyses did not include restrictions on the imaginary part of the potential.

%From a microscopic point of view, the effective pro\-jec\-ti\-le-tar\-get interaction is the result of the nucleon-nucleon interaction between the nucleons in projectile and target based on Feshbach projection theory~\cite{Fes58,Fes62}. This means the effective interaction does not favor the number of degrees of freedom, i.e., observing breakup. This explains the reason of no significant increase of imaginary part of OMP when incident energy below the Coulomb barrier. 

The analysis presented here shows that, within properly quantified uncertainties and with physically motivated constraints, the usual  near threshold behavior (TA) of the OMP is found to be consistent with elastic scattering data for the $^6$Li~+~$^{209}$Bi and $^6$He~+~$^{208}$Pb reactions. This finding is in contrast to the BTA found in earlier works, which was explained in Refs.~\cite{Hussein06,Canto15} by the influence of the breakup reaction. According to this explanation, the imaginary part of OMP increases continuously with decreasing energy below the barrier due to the assumption that the coupling to the breakup channels in these systems continues to be important even at energies below the barrier. However, such explanation fails to account for the fact that the large breakup coupling effects may cause a large suppression in the fusion cross section~\cite{Canto06}. The imaginary potential accounts for the channels leaving the ground states, which include breakup and fusion channels. Therefore, we expect no significant increase in the imaginary potential at energies below the Coulomb barrier. In general, we expect the potential energy dependence to be an universal phenomenon within the barrier-energy region. In other words, even for weakly-bound exotic system, we expect the energy dependence of the optical potential to be similar to the one in tightly-bound systems.

\section{Summary and conclusions}
We have studied the problem of near-threshold potential behavior of weakly-bound nucleus induced reactions using $^6$Li+$^{209}$Bi and $^6$He+$^{208}$Pb as case studies. We have implemented a new method to extract the OMP parameters. The extracted potentials can reproduce the experimental data with $\chi^2/{\rm d.o.f.}\simeq 1$. 

%For the energy below the Coulomb barrier region, the nuclear forces is mostly hidden by the Coulomb interaction resulting in large experimental uncertainties. Rigorous uncertainty quantification with dispersive relation are needed to extract the OMP from the data.   

%By combining a rigorous uncertainty quantification of elastic scattering data with a physical constraint of the imaginary part of the OMP decreasing when the collision energy goes below the Coulomb barrier, a usual near-threshold behaviour (TA) on the real part of OMP is found for weakly bound projectile induced reactions. Contrary to the findings in the other works, we show that experimental data is consistent with the usual TA.

For the energy below the Coulomb barrier region, the nuclear force is mostly hidden by the Coulomb interaction resulting in large experimental uncertainties. Rigorous uncertainty quantification of elastic scattering data with a physical constraint of the imaginary part of the OMP decreasing when the collision energy goes below the Coulomb barrier are needed to extract the OMP from the data. By using these, the real part exhibits the usual near-threshold behavior (TA) in weakly-bound projectile induced reactions contrary to the findings in the other works. In addition we show that the extracted optical potential verifies the dispersive relation. 

Although the studies presented here have been restricted to the $^6$Li and $^6$He projectiles, we believe that the conclusions can be directly extrapolated to other weakly-bound nuclei, such as $^7$Li or $^9$Be. A detailed study with different projectile with various targets is in progress.

%----------------------
\section*{Acknowledgments}
%----------------------
We  are  grateful  to Daniel Phillips, Antonio Moro, Charlotte Elster, and Enrique Ruiz Arriola  for  a  critical  reading  of  the manuscript and many insightful comments on this work.
This work has been partially supported by the National Science Foundation under contract. No.\ NSF-PHY-1520972 with Ohio University, and of the U.S. Department of Energy under Contract No. DE-FG02-93ER40756 with Ohio University.
This research used resources of the National Energy Research Scientific Computing Center (NERSC), a U.S. Department of Energy Office of Science User Facility operated under Contract No. DE-AC02-05CH11231.




%\bibliographystyle{elsarticle-harv} 
\bibliographystyle{elsarticle-num}
\bibliography{uncertainty}

\begin{thebibliography}{10}
\expandafter\ifx\csname url\endcsname\relax
  \def\url#1{\texttt{#1}}\fi
\expandafter\ifx\csname urlprefix\endcsname\relax\def\urlprefix{URL }\fi
\expandafter\ifx\csname href\endcsname\relax
  \def\href#1#2{#2} \def\path#1{#1}\fi

\bibitem{Fes58}
H.~Feshbach, A unified theory of nuclear reactions. {I}, Annals of Physics
  5~(2) (1958) 357.

\bibitem{Fes62}
H.~Feshbach, A unified theory of nuclear reactions. {II}, Annals of Physics
  19~(2) (1962) 287--313.

\bibitem{Cornwall1962}
J.~M. Cornwall, M.~A. Ruderman, Mandelstam representation and {Regge} poles
  with absorptive energy-dependent potentials, Phys. Rev. 128 (1962)
  1474--1484.

\bibitem{LIPPERHEIDE196865}
R.~Lipperheide, A.~Schmidt, Energy dependence of phenomenological optical-model
  potentials, Nuclear Physics A 112~(1) (1968) 65 -- 75.

\bibitem{Baeza84}
A.~Baeza, B.~Bilwes, R.~Bilwes, J.~Díaz, J.~Ferrero, Energy-dependent
  renormalization coefficients of folding-model description of
  $^{32}${S}+$^{40}${Ca} elastic scattering, Nuclear Physics A 419~(2) (1984)
  412 -- 428.

\bibitem{Lilley85}
J.~Lilley, B.~Fulton, M.~Nagarajan, I.~Thompson, D.~Banes, Evidence for a
  progressive failure of the double folding model at energies approaching the
  {Coulomb} barrier, Physics Letters B 151~(3) (1985) 181 -- 184.

\bibitem{Fulton85}
B.~Fulton, D.~Banes, J.~Lilley, M.~Nagarajan, I.~Thompson, Energy dependence of
  the $^{16}${O} + $^{60}${Ni} potential and the optical model dispersion
  relation, Physics Letters B 162~(1) (1985) 55 -- 58.

\bibitem{Nagarajan85}
M.~A. Nagarajan, C.~C. Mahaux, G.~R. Satchler, Dispersion relation and the
  low-energy behavior of the heavy-ion optical potential, Phys. Rev. Lett. 54
  (1985) 1136--1138.

\bibitem{Satchler91}
G.~Satchler, Heavy-ion scattering and reactions near the {Coulomb} barrier and
  “threshold anomalies”, Physics Reports 199~(3) (1991) 147 -- 190.

\bibitem{Mahaux86}
C.~Mahaux, H.~Ngô, G.~Satchler, Causality and the threshold anomaly of the
  nucleus-nucleus potential, Nuclear Physics A 449~(2) (1986) 354 -- 394.

\bibitem{Hussein06}
M.~S. Hussein, P.~R.~S. Gomes, J.~Lubian, L.~C. Chamon, New manifestation of
  the dispersion relation: Breakup threshold anomaly, Phys. Rev. C 73 (2006)
  044610.

\bibitem{li6al27}
J.~M. Figueira, J.~O.~F. Niello, D.~Abriola, A.~Arazi, O.~A. Capurro, E.~d.
  Barbar\'a, G.~V. Mart\'{\i}, D.~M. Heimann, A.~E. Negri, A.~J. Pacheco,
  I.~Padr\'on, P.~R.~S. Gomes, J.~Lubian, T.~Correa, B.~Paes, Breakup threshold
  anomaly in the elastic scattering of $^{6}${Li} on $^{27}${Al}, Phys. Rev. C
  75 (2007) 017602.

\bibitem{li6si28}
A.~Pakou, N.~Alamanos, G.~Doukelis, A.~Gillibert, G.~Kalyva, M.~Kokkoris,
  S.~Kossionides, A.~Lagoyannis, A.~Musumarra, C.~Papachristodoulou,
  N.~Patronis, G.~Perdikakis, D.~Pierroutsakou, E.~C. Pollacco, K.~Rusek,
  Elastic scattering of $^{7}${Li}+$^{28}${Si} at near-barrier energies, Phys.
  Rev. C 69 (2004) 054602.

\bibitem{li6ni58}
M.~Biswas, S.~Roy, M.~Sinha, M.~Pradhan, A.~Mukherjee, P.~Basu, H.~Majumdar,
  K.~Ramachandran, A.~Shrivastava, The study of threshold behaviour of
  effective potential for $^{6}${Li}+$^{58,64}${Ni} systems, Nuclear Physics A
  802~(1) (2008) 67 -- 81.

\bibitem{li6co59}
F.~A. Souza, L.~A.~S. Leal, N.~Carlin, M.~G. Munhoz, R.~L. Neto, M.~M.~d.
  Moura, A.~A.~P. Suaide, E.~M. Szanto, A.~S.~d. Toledo, J.~Takahashi, Effect
  of breakup on elastic scattering for the $^{6,7}${Li}+$^{59}${Co} systems,
  Phys. Rev. C 75 (2007) 044601.

\bibitem{li6zn64}
M.~Zadro, P.~Figuera, A.~D. Pietro, F.~Amorini, M.~Fisichella, O.~Goryunov,
  M.~Lattuada, C.~Maiolino, A.~Musumarra, V.~Ostashko, M.~Papa, M.~G.
  Pellegriti, F.~Rizzo, D.~Santonocito, V.~Scuderi, D.~Torresi, Elastic
  scattering of $^{6}\mathrm{Li}$ on $^{64}\mathrm{Zn}$ at near-barrier
  energies, Phys. Rev. C 80 (2009) 064610.

\bibitem{li6se80}
L.~Fimiani, J.~M. Figueira, G.~V. Mart\'{\i}, J.~E. Testoni, A.~J. Pacheco,
  W.~H.~Z. C\'ardenas, A.~Arazi, O.~A. Capurro, M.~A. Cardona, P.~Carnelli,
  E.~de~Barbar\'a, D.~Hojman, D.~Martinez~Heimann, A.~E. Negri, Elastic
  scattering in the $^{6,7}${Li} $+$ $^{80}${Se} systems, Phys. Rev. C 86
  (2012) 044607.

\bibitem{li6zr90}
H.~Kumawat, V.~Jha, B.~J. Roy, V.~V. Parkar, S.~Santra, V.~Kumar, D.~Dutta,
  P.~Shukla, L.~M. Pant, A.~K. Mohanty, R.~K. Choudhury, S.~Kailas, Breakup
  threshold anomaly in the elastic scattering for the $^{6}${Li}+$^{90}${Zr}
  system, Phys. Rev. C 78 (2008) 044617.

\bibitem{li6sn112}
N.~N. Deshmukh, S.~Mukherjee, D.~Patel, N.~L. Singh, P.~K. Rath, B.~K. Nayak,
  D.~C. Biswas, S.~Santra, E.~T. Mirgule, L.~S. Danu, Y.~K. Gupta, A.~Saxena,
  R.~K. Choudhury, R.~Kumar, J.~Lubian, C.~C. Lopes, E.~N. Cardozo, P.~R.~S.
  Gomes, Breakup threshold anomaly in the near-barrier elastic scattering of
  $^{6}${Li}+$^{116,112}${Sn}, Phys. Rev. C 83 (2011) 024607.

\bibitem{li6ba138}
A.~M.~M. Maciel, P.~R.~S. Gomes, J.~Lubian, R.~M. Anjos, R.~Cabezas, G.~M.
  Santos, C.~Muri, S.~B. Moraes, R.~Liguori~Neto, N.~Added, N.~Carlin~Filho,
  C.~Tenreiro, Influence of the ${}^{6,7}\mathrm{Li}$ breakup process on the
  near barrier elastic scattering by heavy nuclei, Phys. Rev. C 59 (1999)
  2103--2107.

\bibitem{li6sm144}
J.~M. Figueira, J.~O.~F. Niello, A.~Arazi, O.~A. Capurro, P.~Carnelli,
  L.~Fimiani, G.~V. Mart\'{\i}, D.~M. Heimann, A.~E. Negri, A.~J. Pacheco,
  J.~Lubian, D.~S. Monteiro, P.~R.~S. Gomes, Energy dependence of the optical
  potential of weakly and tightly bound nuclei as projectiles on a medium-mass
  target, Phys. Rev. C 81 (2010) 024613.

\bibitem{li6pb208}
N.~Keeley, S.~Bennett, N.~Clarke, B.~Fulton, G.~Tungate, P.~Drumm,
  M.~Nagarajan, J.~Lilley, Optical model analyses of $^{6,7}${Li} +
  $^{208}${Pb} elastic scattering near the {Coulomb} barrier, Nuclear Physics A
  571~(2) (1994) 326 -- 336.

\bibitem{li6bi209}
S.~Santra, S.~Kailas, K.~Ramachandran, V.~V. Parkar, V.~Jha, B.~J. Roy,
  P.~Shukla, {Reaction mechanisms involving weakly bound $^{6}${Li} and
  $^{209}${Bi} at energies near the {Coulomb} barrier}, Phys. Rev. C83 (2011)
  034616.

\bibitem{li6th232}
S.~Dubey, S.~Mukherjee, D.~C. Biswas, B.~K. Nayak, D.~Patel, G.~K. Prajapati,
  Y.~K. Gupta, B.~N. Joshi, L.~S. Danu, S.~Mukhopadhyay, B.~V. John, V.~V.
  Desai, S.~V. Suryanarayana, R.~P. Vind, N.~N. Deshmukh, S.~Appnnababu, P.~M.
  Prajapati, Effect of breakup processes on the near-barrier elastic scattering
  of the $^{6,7}${Li} + $^{232}${Th} systems, Phys. Rev. C 89 (2014) 014610.

\bibitem{li7al27}
J.~M. Figueira, D.~Abriola, J.~O.~F. Niello, A.~Arazi, O.~A. Capurro, E.~d.
  Barbar\'a, G.~V. Mart\'{\i}, D.~Mart\'{\i}nez~Heimann, A.~J. Pacheco, J.~E.
  Testoni, I.~Padr\'on, P.~R.~S. Gomes, J.~Lubian, Absence of the threshold
  anomaly in the elastic scattering of the weakly bound projectile $^{7}${Li}
  on $^{27}${Al}, Phys. Rev. C 73 (2006) 054603.

\bibitem{li7Zn64}
M.~M. Shaikh, M.~Das, S.~Roy, M.~Sinha, M.~Pradhan, P.~Basu, U.~Datta,
  K.~Ramachandran, A.~Shrivastava, Threshold behavior of interaction potential
  for the system $^{7}${Li} + $^{64}${Ni}: Comparison with $^{6}${Li} +
  $^{64}${Ni}, Nuclear Physics A 953 (2016) 80 -- 94.

\bibitem{li7pb208}
L.~Yang, C.~J. Lin, H.~M. Jia, F.~Yang, Z.~D. Wu, X.~X. Xu, H.~Q. Zhang, Z.~H.
  Liu, P.~F. Bao, L.~J. Sun, N.~R. Ma, Optical model potentials for the
  $^{6}${He}+$^{209}${Bi} reaction from a
  $^{208}\mathrm{Pb}$($^{7}\mathrm{Li}$,$^{6}\mathrm{He}$)$^{209}\mathrm{Bi}$
  reaction analysis, Phys. Rev. C 89 (2014) 044615.

\bibitem{be9al27}
P.~R.~S. Gomes, R.~M. Anjos, C.~Muri, J.~Lubian, I.~Padron, L.~C. Chamon, R.~L.
  Neto, N.~Added, J.~O. Fern\'andez~Niello, G.~V. Mart\'{\i}, O.~A. Capurro,
  A.~J. Pacheco, J.~E. Testoni, D.~Abriola, Threshold anomaly with weakly bound
  projectiles: Elastic scattering of $^{9}${Be}+$^{27}${Al}, Phys. Rev. C 70
  (2004) 054605.

\bibitem{be9y89}
C.~S. Palshetkar, S.~Santra, A.~Shrivastava, A.~Chatterjee, S.~K. Pandit,
  K.~Ramachandran, V.~V. Parkar, V.~Nanal, V.~Jha, B.~J. Roy, S.~Kalias,
  Elastic scattering and $\ensuremath{\alpha}$ production in the
  $^{9}${Be}+$^{89}${Y} system, Phys. Rev. C 89 (2014) 064610.

\bibitem{be9sm144}
P.~Gomes, J.~Lubian, B.~Paes, V.~Garcia, D.~Monteiro, I.~Padrón, J.~Figueira,
  A.~Arazi, O.~Capurro, L.~Fimiani, A.~Negri, G.~Martí, J.~F. Niello,
  A.~Gómez-Camacho, L.~Canto, Near-barrier fusion, breakup and scattering for
  the $^{9}${Be}+$^{144}${Sm} system, Nuclear Physics A 828~(3) (2009) 233 --
  252.

\bibitem{be9pb208}
R.~J. Woolliscroft, B.~R. Fulton, R.~L. Cowin, M.~Dasgupta, D.~J. Hinde, C.~R.
  Morton, A.~C. Berriman, Elastic scattering and fusion of
  $^{9}\mathrm{Be}+^{208}\mathrm{Pb}$: Density function dependence of the
  double folding renormalization, Phys. Rev. C 69 (2004) 044612.

\bibitem{Yang17}
L.~Yang, C.~J. Lin, H.~M. Jia, D.~X. Wang, N.~R. Ma, L.~J. Sun, F.~Yang, X.~X.
  Xu, Z.~D. Wu, H.~Q. Zhang, Z.~H. Liu, Is the dispersion relation applicable
  for exotic nuclear systems? {The} abnormal threshold anomaly in the
  $^{6}${He}+$^{209}${Bi} system, Phys. Rev. Lett. 119 (2017) 042503.

\bibitem{yang17prc}
L.~Yang, C.~J. Lin, H.~M. Jia, D.~X. Wang, N.~R. Ma, L.~J. Sun, F.~Yang, X.~X.
  Xu, Z.~D. Wu, H.~Q. Zhang, Z.~H. Liu, Abnormal behavior of the optical
  potential for the halo nuclear system $^{6}${He}+$^{209}${Bi}, Phys. Rev. C
  96 (2017) 044615.

\bibitem{james1972}
F.~James, MINUIT - Function Minimization and Error Analysis, CERN Program
  Library Long Writeup D506, CERN, 1972.

\bibitem{SanchezBenitez:2005jj}
A.~M. Sanchez-Benitez, et~al., {Scattering of $^{6}${He} at energies around the
  {Coulomb} barrier}, J. Phys. G31 (2005) S1953--S1958.

\bibitem{Sanchez-Benitez:2008krm}
A.~M. Sánchez-Benítez, et~al., {Study of the elastic scattering of $^{6}${He}
  on $^{208}${Pb} at energies around the {Coulomb} barrier}, Nucl. Phys. A803
  (2008) 30--45.

\bibitem{Marquinez-Duran:2012ilt}
G.~Marquínez-Durán, et~al., {Scattering of $^{8}${He} on $^{208}${Pb} at
  Energies Around the {Coulomb} Barrier}, Acta Phys. Polon. B43~(2) (2012) 239.

\bibitem{Kakuee:2005kq}
O.~R. Kakuee, et~al., {Long range absorption in the scattering of $^{6}$He on
  $^{208}$Pb and $^{197}$Au at 27 MeV}, Nucl. Phys. A765 (2006) 294--306.

\bibitem{thompson2009nuclear}
I.~Thompson, F.~Nunes, Nuclear Reactions for Astrophysics: Principles,
  Calculation and Applications of Low-Energy Reactions, Cambridge University
  Press, 2009.

\bibitem{fresco}
I.~J. Thompson, Coupled reaction channels calculations in nuclear physics,
  Computer Physics Reports 7~(4) (1988) 167 -- 212.

\bibitem{press1992numerical}
W.~Press, S.~Teukolsky, W.~Vetterling, B.~Flannery, Numerical Recipes in
  FORTRAN: The Art of Scientific Computing, Vol. 1-2, Cambridge University
  Press, 1992.

\bibitem{Perez:2013jpa}
R.~Navarro~Pérez, J.~E. Amaro, E.~Ruiz~Arriola, {Coarse-grained potential
  analysis of neutron-proton and proton-proton scattering below the pion
  production threshold}, Phys. Rev. C88~(6) (2013) 064002, [Erratum: Phys.
  Rev.C91,no.2,029901(2015)].

\bibitem{Perez:2015pea}
R.~Navarro~Pérez, E.~Ruiz~Arriola, J.~Ruiz~de Elvira, {Self-consistent
  statistical error analysis of $\pi\pi$ scattering}, Phys. Rev. D91 (2015)
  074014.

\bibitem{Furnstahl:2014xsa}
R.~J. Furnstahl, D.~R. Phillips, S.~Wesolowski, {A recipe for EFT uncertainty
  quantification in nuclear physics}, J. Phys. G42~(3) (2015) 034028.

\bibitem{Perez:2014kpa}
R.~Navarro~Pérez, J.~E. Amaro, E.~Ruiz~Arriola, {Error analysis of nuclear
  forces and effective interactions}, J. Phys. G42~(3) (2015) 034013.

\bibitem{Pastore:2018xuu}
A.~Pastore, An introduction to bootstrap for nuclear physics, J. Phys. G.
  46~(5)  052001.

\bibitem{Navarro:2014laa}
R.~Navarro~Pérez, E.~Garrido, J.~E. Amaro, E.~Ruiz~Arriola, {Triton binding
  energy with realistic statistical uncertainties}, Phys. Rev. C90~(4) (2014)
  047001.

\bibitem{Perez:2014jsa}
R.~Navarro~Pérez, J.~E. Amaro, E.~Ruiz~Arriola, {Bootstrapping the statistical
  uncertainties of NN scattering data}, Phys. Lett. B738 (2014) 155--159.

\bibitem{Perez:2015bqa}
R.~Navarro~Pérez, J.~E. Amaro, E.~Ruiz~Arriola, P.~Maris, J.~P. Vary,
  {Statistical error propagation in ab initio no-core full configuration
  calculations of light nuclei}, Phys. Rev. C92~(6) (2015) 064003.

\bibitem{Pasquini:2017ehj}
B.~Pasquini, P.~Pedroni, S.~Sconfietti, {First extraction of the scalar proton
  dynamical polarizabilities from real Compton scattering data}, Phys. Rev.
  C98~(1) (2018) 015204.

\bibitem{Abriola:2017}
{Abriola, Daniel}, {Mart\'{\i}, Guillermo V.}, {Testoni, Jorge E.}, Bootstrap
  method for constructing covariance matrices of optical-model parameters in
  the study of the threshold anomaly, EPJ Web Conf. 146 (2017) 02050.

\bibitem{Abriola:2015ufa}
D.~Abriola, A.~Arazi, J.~Testoni, F.~Gollan, G.~V. Martí, {Uncertainties of
  optical-model parameters for the study of the threshold anomaly}, J. Phys.
  Conf. Ser. 630~(1) (2015) 012021.

\bibitem{Canto15}
L.~Canto, P.~Gomes, R.~Donangelo, J.~Lubian, M.~Hussein, Recent developments in
  fusion and direct reactions with weakly bound nuclei, Physics Reports 596
  (2015) 1 -- 86, recent developments in fusion and direct reactions with
  weakly bound nuclei.

\bibitem{Canto06}
L.~Canto, P.~Gomes, R.~Donangelo, M.~Hussein, Fusion and breakup of weakly
  bound nuclei, Physics Reports 424~(1) (2006) 1 -- 111.

\end{thebibliography}

%\begin{thebibliography}{00}
%
%%% \bibitem[Author(year)]{label}
%%% Text of bibliographic item
%\end{thebibliography}
\end{document}